\documentclass[pageno]{jpaper}

\newcommand{\asplossubmissionnumber}{346}

\usepackage[normalem]{ulem}
\usepackage{pifont}
\usepackage{array}
\usepackage{multirow}

\begin{document}

\title{GMEM: Generalized Memory Management for Peripheral Devices}

\author{
{\rm Weixi Zhu, Alan L. Cox and Scott Rixner}\\
{\rm Rice University} \\
{\rm \{wxzhu, alc, rixner\}@rice.edu}
} 

\date{}
\maketitle

\begin{abstract}


This paper presents GMEM, generalized memory management, for
peripheral devices. GMEM provides OS support for centralized memory
management of both CPU and devices.
GMEM provides a high-level interface that decouples
MMU-specific functions. Device drivers can thus attach themselves to
a process's address space and let the OS take charge of their
memory management. This eliminates the need for device drivers to
``reinvent the wheel" and allows them to benefit from general 
memory optimizations integrated by GMEM.
Furthermore, GMEM internally
coordinates all attached devices within each virtual address space. This
drastically improves user-level programmability, since
programmers can use a single address space within their program,
even when operating across the CPU and multiple devices.
A case study on device drivers demonstrates these benefits.
A GMEM-based IOMMU driver eliminates around seven hundred lines
of code and obtains 54\% higher network receive throughput utilizing
32\% less CPU compared to the state-of-the-art. In addition,
the GMEM-based driver of a simulated GPU takes less than 70 lines of 
code, excluding its MMU functions.

\end{abstract}

\section{Introduction}

In the quest for increased performance~\cite{esmaeilzadeh2011dark,
hennessy2019new}, there will continue to be a rise of specialized
hardware accelerators~\cite{dally2020domain, turakhia2018darwin,
chen2014diannao, jouppi2017datacenter, suda2016throughput,
han2016eie}, increased integration with GPUs, networking, and
storage, and large amounts of non-volatile memory (NVM) across the system.
This conglomeration of components in next generation computing
systems requires rethinking the ways in which disparate memory is
managed. Realizing the full potential of these systems will require
significant innovations in systems
software~\cite{ausavarungnirun2017mosaic, shan2018legoos,
yan2019nimble, agarwal2017thermostat, korolija2020abstractions,
khawaja2018sharing}.

Unfortunately, modern operating system (OS) virtual memory management
systems have no high-level kernel programming interfaces (KPIs) for
peripheral devices to easily leverage existing virtual memory (VM)
mechanisms. A heterogeneous system no longer solely
manages address mappings for processes running on the CPU. Instead,
it now must also manage the address spaces for virtualization, I/O
devices, accelerators, GPUs, NVM, and other components of the system.
To date, these diverse peripheral devices have been
managed with independent, disconnected systems, leading to a
proliferation of virtual address space management, physical memory
management, and address mapping systems within the OS.  

It is costly to implement and maintain an independent memory 
management system for a device.
As each new device is added to the system, the device and OS
developers have been ``reinventing the wheel'' to build address space
managers, memory allocators, and mapping managers unique to the
device.  These systems are often optimized for the unique
characteristics of the device, but are still accomplishing the same
fundamental tasks.  And inevitably, they suffer from missing features
or missing performance optimizations that exist in the core VM system, largely
due to the difficulty of replicating the complex mechanisms that have
evolved over time in the core VM system.

Independent peripheral memory management also presents new 
coordination challenges.
Peripheral devices can have unique page table formats and unique
synchronization mechanisms~\cite{gp100_mmu, power2014supporting,
amdIOMMU}, all of which have to be supported simultaneously.  When an
address space is shared between the CPU and one or more devices (or
just among devices), these disparate page tables must be kept
coherent.  I/O devices also have a much higher churn rate on address
mappings than the CPU in order to maintain security and isolation
from the device~\cite{markettos2019thunderclap, morgan2018iommu,
ben2007price}.  Furthermore, many peripheral devices have their own
local physical memory.  This memory must be managed across the system
and the VM system must determine whether to migrate data between
device memory and host memory or to force the use of remote
accesses~\cite{li2019framework}.

Ultimately, these challenges make somewhat simple-seeming operating
system tasks extremely complicated.  For example, if the operating
system wants to destroy all of the mappings to a specific physical
page in memory, it now potentially has to be aware of every subsystem
that might hold a mapping to that page and destroy each mapping in a
device-specific way.  The state-of-the-art is to use notifiers to
attempt to simplify this process.  For example, Linux's MMU notifier
mechanism can be used to notify a secondary VM system that a shared
page has been invalidated by the core VM system~\cite{mmu_notifier}.
While this mechanism facilitates coordination between a secondary
and the core VM system, it does not eliminate the need for a
secondary VM system in the device driver.
Meanwhile, many code maintenance issues are
introduced between the OS and device
drivers~\cite{last_minute_notifier, linus_complains_mmu_notifiers,
hmm_mmu_notifier}.

This paper proposes \textit{G}eneralized \textit{ME}mory
\textit{M}anagment (\textit{GMEM}) for peripheral devices. GMEM
provides centralized memory management for systems utilizing heterogeneous
memory resources. Specifically, GMEM introduces three key
innovations.  First, GMEM provides a high-level interface within the
operating system for the management of memory across all devices.
This enables devices to leverage the operating system's
memory management subsystem effectively.  Each device driver only needs to implement the device-specific code that is
necessary to enable the operating system to interact with the
hardware, rather than inventing a completely new memory management system
for every device.  Second, GMEM leverages the functionality and
optimizations within the VM system across devices.  Therefore, 
as the OS integrates additional capabilities and performance
optimizations, GMEM-based device drivers automatically benefit.
Finally, GMEM enables the use of a single address space across
the CPU and devices.  This improves the programmability of
accelerators, as GMEM's internal coordination mechanisms perform
the necessary memory management tasks, freeing the programmer
from worrying about address space coordination.


The effectiveness of the GMEM system is demonstrated using several
case studies.  For example, converting FreeBSD's IOMMU driver to
use GMEM's KPI eliminates the need for a special-purpose IOVA allocator. This
eliminates over six hundred lines of code and reduces the chance of
performance bugs~\cite{iova_rbbug}. Furthermore, as GMEM
integrates separate memory optimizations~\cite{vmem, brett_thesis}, the
GMEM-based IOMMU driver receives automatic performance improvements.
Experiments show that a GMEM-based driver obtains 54\% higher network receive
throughput while utilizing 32\% less CPU than mainline Linux.
Simultaneously, the GMEM-based IOMMU driver enables line rate
transmit throughput with 45\% less CPU utilization than Linux. As another example, the Intel GPU driver was modified to
use GMEM. GMEM integrates a memory optimization that speeds up zeroing
larger pages~\cite{zhu2020comprehensive}, which reduces the GPU kernel launch time
by 45\%.

GMEM can also simplify driver development.  The drivers of two simulated
GPUs take less than 70 lines of code to implement their device memory
management, excluding the memory management unit (MMU) functions. Applications can quickly be
developed for these simulated devices within a single address space
and they can execute seamlessly with smaller device memory capacity.

These case studies demonstrate that GMEM enables device drivers to
take better advantage of existing and future devices, provides more
functionality, and improves performance. More importantly, the OS
now has centralized control of heterogeneous memory resources
enabled by GMEM's internal coordination. This lays the foundation for
better global resource utilization in systems incorporating multiple
types of accelerators.


    




The rest of the paper proceeds as follows.
Section~\ref{sec:background} introduces the background and related
work of virtual memory management. Section~\ref{sec:design} explains
the design of GMEM. Section~\ref{sec:example} then introduces the GMEM
KPI with two use cases, followed by Section~\ref{sec:impl} that
explains the implementation details. After that,
Section~\ref{sec:case-studies} evaluates GMEM's impact with both
real-world and simulated drivers. Finally,
Section~\ref{sec:conclusion} concludes the paper and discusses future
work.

\section{Background and Related Work}
\label{sec:background}
\subsection{Virtual Memory Management}

The VM system of the OS decouples
addresses used in programs from actual physical memory.  It does so
with the aid of the hardware MMU.  In order to provide the virtual
memory abstraction, the virtual memory system performs four key
operations: virtual address management, logical address mapping,
physical address mapping, and physical memory management.

\paragraph{Virtual Address Management.}

When a new process starts, the VM system creates a new virtual address (VA)
space for that process and sets up this initial VA space.
To allocate additional addresses or change the existing allocations,
the program must make a system call (such as {\tt mmap()}, {\tt
  munmap()}, {\tt sbrk()}, {\tt exec()}, etc.).  These system calls
allocate, change, or deallocate the virtual addresses that can then be
used by the program.  Some of these system calls may also include
declarations or changes to the memory access privileges of those
address ranges, as well.  Note that no physical memory is actually
allocated when virtual addresses are allocated.

\paragraph{Logical Address Mapping.}

As previously stated, when a virtual address range is allocated by a
system call, no
physical memory is actually allocated.  Instead, a logical mapping is
created that defines where the initial values will come from when (and
if) the physical memory is allocated to back this virtual address
range.  For example, the physical memory may be filled with the
contents of a file, or simply zeroes.

\paragraph{Physical Address Mapping.}

The VM system provides virtual-to-physical mapping management to
control the address translation process of a CPU's MMU.  The page table stores the existing, valid
mappings from virtual addresses to physical addresses. The hardware MMU
uses the page table to translate addresses and caches those
translations in the translation lookaside buffer (TLB).  If there is
no valid translation in the page table (or the accessor does not have
the appropriate privileges to use the translation), the hardware will
cause a page fault that can then be handled by the VM system.

Physical mappings are not created eagerly when virtual addresses are
allocated.
Instead, physical mappings are created on first access.  When a program
accesses an allocated virtual address that has not yet been physically mapped,
the page fault handler will create a virtual-to-physical address
mapping. If physical memory has already been allocated and prepared to back that
virtual address, the mapping can be created and installed in the page
table immediately.  Otherwise, physical memory must first be allocated
and prepared (to be discussed next).  When the process is resumed, the
translation will be re-attempted by the hardware MMU and the physical memory
can then be accessed. 
Note that if a process accesses an unallocated virtual address or does
not have the appropriate privileges for the type of access, the page
fault handler will instead terminate the program.

Mappings are created at page size granularities.  X86-64 processors
support 4KB, 2MB, and 1GB page sizes.  The VM system is responsible
for selecting the page size and potentially changing it in response
to various events, for example, ``transparent huge pages'' in Linux
or ``automatic superpage promotion'' in FreeBSD.

When virtual addresses are deallocated, their associated mappings must
also be destroyed.  This means removing the mappings from the page
table(s) and invalidating the mappings that are cached in any hardware
MMUs.

\paragraph{Physical Memory Management.}

The VM system does not allocate physical memory until an allocated
virtual address is accessed and needs to be physically mapped to new physical
memory.  On a page fault, if the virtual address was valid and
physical memory has not already been allocated, the VM system
allocates physical memory before a virtual-to-physical address mapping
can be created.

Physical memory can be allocated a single page at a time (using any of
the valid page sizes) or multiple contiguous pages could be allocated
at once.  After physical memory has been allocated, it must be
``prepared''.  This potentially involves zero-filling the page (if the
page was not already cleared), copying memory (to support
copy-on-write, for example), or reading the contents from a file on
disk. The logical mapping for the virtual address that is being backed
by this physical memory indicates how the physical memory should be
prepared.

If there is no free physical memory, then the VM system must
``reclaim'' physical memory for use.  In order to reclaim a page of physical
memory, it may first need to be stored to disk so that it can later be
recovered when it is next accessed.  If the page contents are already on disk
or is otherwise unneeded, the page can be reclaimed immediately.

Once physical memory has been allocated and prepared, a
virtual-to-physical address mapping can be created, as described
above.

\subsection{Peripheral memory management}


\begin{table}[t]
\centering
\begin{tabular}{|c|cc|}
\hline
\multirow{2}{*}{\begin{tabular}[c]{@{}c@{}}Physical\\ Memory\end{tabular}} & \multicolumn{2}{c|}{VA Space}                          \\ \cline{2-3} 
                                                                           &
\multicolumn{1}{c|}{Private}    & Shared               \\ \hline
Local
& \multicolumn{1}{c|}{CUDA} & UVM            \\ \hline Shared
& \multicolumn{1}{c|}{BUS\_DMA} & KVM pass-through \\ \hline
\end{tabular}
\caption{Peripheral VM use cases.}
\label{table:va_cases}
\end{table}


Historically, devices have been given direct, unrestricted access to a
machine's memory, which was entirely controlled by
trusted kernel code. Moreover, any firmware within the device was
trusted. However, today many devices are controlled by untrusted
third-party drivers that can cause security issues. Therefore,
peripheral users may isolate their memory resources with VA spaces to
restrict access from other devices. 
Table~\ref{table:va_cases} categorizes these use cases into four 
examples, depending on whether the VA space is
shared with a host process and whether the device has its own local
physical memory.




The BUS\_DMA interface supports a per-device private VA space to
isolate each device to access only permitted machine memory within
its own VA space. Similarly, Nvidia's CUDA programming model
supports a per-GPU private VA space to protect data stored in the GPU's local physical memory.

There have also been efforts to support shared VA spaces,
including OpenCL's shared virtual memory (SVM) and Nvidia's
unified virtual memory (UVM) systems.  UVM, for example, unifies
the GPU and CPU address spaces so that UVM programs do not need
to manage device buffers or orchestrate transfers across VA
spaces.  KVM pass-through devices share both the VA space
and the physical memory between the guest virtual machine and
the device.

The proposed GMEM system provides a generalized high-level interface for
device drivers to support all possible use cases as exemplified in
Table~\ref{table:va_cases}. In contrast, existing OS support provides
very limited support, as elaborated below.

\subsubsection{No prior work helps per-device private VA space}




The BUS\_DMA interfaces from FreeBSD and Linux both implement
binary-search trees to manage I/O virtual addresses in private VA spaces.
The core OS VM systems already implement such a system for managing
process address spaces on the CPU, but the device drivers cannot
easily reuse those core systems.

This is not an isolated problem. In fact, most memory management tasks for
peripherals are no different from those for the CPU. However, the
core VM system does not expose high-level KPIs for peripheral drivers
to easily access the core mechanisms. Reimplementing these mechanisms
is complex and error prone -- the FreeBSD VM system has over 30K LoC and
the Linux VM system has more than 80K LoC. As a real-world example, Nvidia's
CUDA driver implements a full GPU memory management system.
It takes at least 11K LoC to implement virtual address management and
logical mapping management, over 6K LoC for physical mapping
management and more than 17K LoC for physical memory management.
While the GPU VM system (a total of over 34K LoC) is a small part of
Nvidia's CUDA driver (695K LoC)~\cite{nvidia_driver}, it is of comparable
size to core OS VM systems.

\subsubsection{Linux helps peripherals share CPU VA spaces}

Linux provides a few low-level mechanisms (MMU notifiers and
heterogeneous memory management) for peripherals to
coordinate with the core VM system when sharing CPU processes'
address spaces~\cite{mmu_notifier, linux_hmm}. However, these low-level
mechanisms do not fully eliminate the need for device drivers to implement
coordination mechanisms. In contrast, GMEM-based drivers that utilize
GMEM's high-level interface can seamlessly share VA spaces, as GMEM
internally handles coordination between the CPU and devices.

Furthermore, Linux's low-level mechanisms are far from sufficient as
a generalized solution for peripherals. Only three core VM events are
supported: destruction of mappings, restriction of mappings, and
CPU access to device private memory.  Moreover, the coordination of these
events are ``one-way'' in that a driver can ask the core VM system to
notify it when these events occur, but the OS or another driver
cannot ask the driver to make similar notifications.

\paragraph{Linux's MMU notifier.}

The peripheral driver can create and insert an MMU notifier on a CPU
process' address space and get driver-specific callback functions
automatically invoked by the core VM system upon certain events, e.g.
when a host physical address mapping is restricted or destroyed.
Drivers may further utilize the MMU interval notifier that restricts
the invocation of the callback functions within specified virtual
address regions~\cite{mmu_interval_notifier}. The MMU interval
notifier has a built-in interval tree that can help track the
logical mappings of the user drivers.

The MMU notifier was originally invented to simplify the
process of hot-plugging a guest
virtual machine's physical memory, where the guest machine must react
by updating the guest page tables. It was later used by the IOMMU
driver to support shared virtual memory (SVM) for peripherals, where
the CPU process's address space is shared by IOMMU-enabled
peripherals for DMA operations.

In Linux 5.13, there are nine drivers that actively use the MMU
notifier. Two of them have no actual peripherals, including Xen and
KVM. The rest of them use it for three purposes --
invalidating a physical address mapping, invalidating a logical
address mapping tracked by the driver, and resubmitting hardware
commands when detecting host physical address mapping invalidation.

\paragraph{Heterogeneous memory management.}

Linux's heterogeneous memory management (HMM) includes several helper
functions, an extended data structure, and a notification
mechanism~\cite{linux_hmm}. They simplify the coordination
tasks that must be performed by peripheral drivers,
including VA space sharing and host-device memory
migration.

To share a CPU process's VA space, the peripheral driver synchronizes
the device's logical and physical mappings using the following steps. The
driver first uses one of HMM's helper functions to inspect CPU
physical mappings. Then, it parses these mappings and establishes the
same logical mappings. According to the logical mappings, physical
mappings can be created in device page tables. At this point, the
device has a coherent view of memory with the CPU process.
The helper function, named \texttt{hmm\_range\_fault}, scans a given
virtual address range and can optionally invoke the CPU page fault
handler to create missing CPU physical mappings.
\texttt{Hmm\_range\_fault} utilizes Linux's MMU notifier to guarantee
that no host physical address mappings are destroyed or restricted
before it returns.

Device drivers can use the extended data structure, \texttt{struct
page}, and HMM's notification mechanism to implement
device-to-host data transfers. 
The \texttt{struct page} can be
utilized to track device physical memory at a 4KB granularity, but
the driver must correctly use HMM's three migration helper functions
when migrating memory between the host and devices. Otherwise, bugs
can be introduced in the core VM system. This is the peril of
exposing one of the core VM's data structures to third-party
drivers.
HMM's notification mechanism invokes the device-to-host migration
code implemented by the device driver when a CPU process accesses
data mapped to device's local physical memory. The CPU page table is
abused to store information that instructs the page fault handler
where the data resides. Currently, only Nouveau and IBM's secure KVM
exploits HMM's one-way notification mechanism in Linux 5.13. In the
opposite direction, the device driver has to use low-level functions
in the core VM system to implement host-to-device migration.

\subsubsection{Limitations against GMEM.}

Linux's MMU notifiers and HMM are low-level mechanisms for
implementing some aspects of peripheral memory management.  They
include low-level KPIs and extended kernel data structures to help
coordinate with the core VM system. This approach has drawbacks.
First, the driver must reinvent its own VM system most of which is
likely hardware-independent. Second, sharing low-level kernel data
structures introduces the potential for catastrophic bugs.
The entire core VM code may crash if a device
driver does not correctly 
use the low-level KPIs to manipulate the kernel data structures.

The biggest drawback of this low-level KPI approach is that it couples
the device drivers and the core VM system in an undesirable way.  Device
drivers can unintentionally misuse the KPIs and all parts of the core
VM system must maintain the new invariants introduced by these low
level mechanisms.  This has led to several subtle issues.
For example, the MMU
notifier's callback function was historically disallowed to sleep,
but drivers want to issue and wait for device commands within the
callback function~\cite{linus_complains_mmu_notifiers}. Updating the
core VM system to allow sleeping callback functions will require
device drivers to update their code as
well~\cite{last_minute_notifiers}. Conversely, the core VM system
must understand and respond correctly to notification
events~\cite{hmm_mmu_notifier}. For example, Linux's out-of-memory
reaper did not invoke notifiers, which could cause situations where
the reaper had reclaimed memory, but a device's driver would not know
this, leading to memory corruption~\cite{last_minute_notifiers}. This
ultimately made device drivers including Intel and AMD's GPU drivers
implement their own coordination mechanisms, where the MMU notifier
was simply used as a synchronization mechanism to minimize the
requirement of code changes from the core VM system.

In contrast, this paper presents GMEM to avoid the above drawbacks. GMEM refactors the
core VM system to provide a centralized memory management system
which supports a machine equipped with various types of accelerators.
No low-level KPIs, extended data structures, or implementation guide
are exposed to the driver developer. Instead, device drivers only
need to encapsulate their hardware-related functions and offload all
the hardware-independent functions to GMEM, using a high-level interface.
GMEM then handles the coordination internally.

\section{Design}
\label{sec:design}

\begin{figure*}[t]
    \centering
    \includegraphics[width=0.95\textwidth]{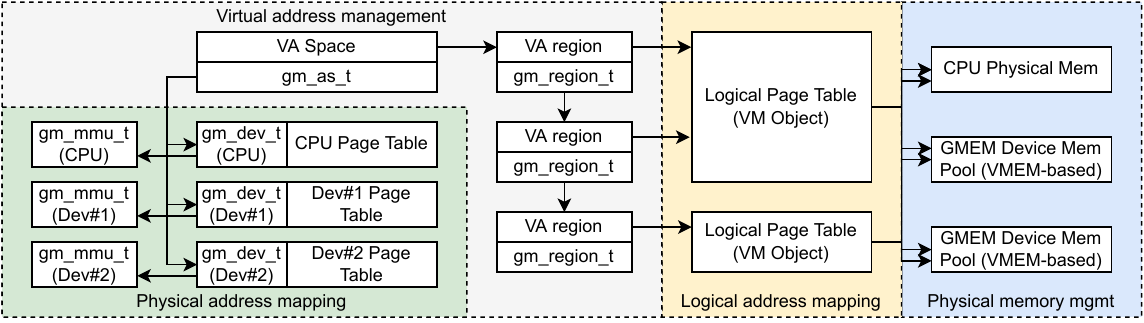}
    \caption{GMEM re-architects the core VM system for peripheral MMUs.}
    \label{fig:gmem-overview}
\end{figure*}

GMEM centers around VA space management and decouples MMU-specific
functions. Device drivers are required to register their own MMU
functions and attach devices to VA spaces, as shown in
Figure~\ref{fig:gmem-overview}. Consequently, GMEM can internally
coordinate the attached devices to provide a coherent view of memory
within each VA space. Device drivers thus do not need to implement any
virtual memory management mechanisms. GMEM's internal coordination
is transparent to device drivers.

GMEM provides two-way coordination between each pair of devices attached
to the same VA space. This is realized with a logical page table that
can map virtual addresses to any local physical memory of the attached
devices. The modified page fault handler can look up the logical page
table to know whether the physical memory has been mapped by one of
the other attached devices. GMEM then decides whether to migrate
memory for local memory accesses or directly create a physical mapping
for remote memory accesses. Both decisions require GMEM to invoke the
registered MMU functions of the two devices. This allows, but does
not require, drivers to direct the page fault handler to switch
between {\em fault-driven migration} and {\em remote access} for
specified virtual address regions.

\subsection{Attaching peripherals to VA spaces}

Device drivers do not need to implement any virtual memory management.
Instead, they just need to implement the MMU functions required by
GMEM, create a GMEM device instance with these MMU functions,
and attach the device instance to a process's address space.
GMEM then takes charge of the VM management of the device, including
invoking its MMU functions to manipulate device page tables or
invalidate device TLB entries. Additional hardware information also
must be registered, including the supported page sizes, the
size of device physical memory, MMU capabilities.

The most important MMU capability is whether the device can recover
from a hardware page fault. If so, the device driver should use GMEM
KPI to handle device page faults. Otherwise, all accessible memory
must be mapped on the device at all times. This means that the host
CPU must wire all memory in the attached virtual address space in
order for the device to have access to it.  The user may decide to
only wire memory that the device may legitimately use. That maintains
the coherence property, but restricts the valid virtual addresses that
the device may access.

\subsection{VA space coherence}

Providing a coherent view of memory can significantly user-level
programmability, because programmers do not need to allocate and
manage separate device buffers. Virtual address space coherence relies
on GMEM's internal coordination within each VA space. Specifically,
GMEM supports two modes of coordination, using shared or coherent page
tables.

The shared page table, as the name suggests, is literally shared among
the CPU/devices of the address space.  The MMUs of all sharers refer
to the exact same page table. In order to use this mode, all sharers
must support the exact same page table format. However, they may use
different TLBs, so GMEM still needs to implement TLB shootdowns by
invoking each device's MMU functions.

A coherent page table, in contrast, is a set of per-device page tables
(using the same or different format) that are kept coherent by GMEM.
Each coherent page table does not necessarily contain entries for all
allocated virtual addresses at all times, rather GMEM ensures that the
device sees a coherent view of memory at all times. Note that the {\em
  logical} address mappings are identical across coherent page tables,
each page table may have different {\em physical} address mappings
(corresponding to those logical address mappings) installed at any
given time.

Devices can be attached to the same virtual address space in different
ways.  For instance, two devices could use a shared page table and a
third device could attach to the same virtual address space with a
coherent page table.  GMEM will automatically keep the page tables
updated and maintain the coherency of their TLBs.

Using a coherent page table brings the flexibility of switching
between {\em fault-driven migration} and {\em remote access}, since a
remote memory access can trigger a second page fault. GMEM's current
design takes the second page fault as a hint for false data locality
and actively migrates memory to the faulted device. However, GMEM can
be further extended to let programmers define the device affinity of
an arbitrary virtual memory region. Instead of migrating memory, GMEM
maps remote physical memory in the device's coherent page table, so
the device performs slower, remote access to such regions of memory.


\section{GMEM Interface and Usage Examples}
\label{sec:example}

This section first describes GMEM's interface, detailing its public
functions and data types. Then, it presents two examples that
illustrate how GMEM's interface is used by device drivers for some
representative peripherals.

\subsection{GMEM KPI}
\label{sec:gmem_kpi}

Table~\ref{tab:datatypes} lists the public data types defined by GMEM
as part of its kernel programming interface (KPI).

\begin{table*}
  \centering
  {\footnotesize
    \begin{tabular}{|m{1.15in}|m{5.15in}|}\hline
      {\bf Data Type} & {\bf Description} \\\hline\hline
  {\tt gm\_as\_t}
  &  Represents a virtual address space, including the data structures for
    representing the allocated ranges of virtual addresses and implementing
    virtual-to-physical mappings, such as page tables. \\\hline
  {\tt gm\_va\_t}
  &  Represents a virtual address. \\\hline
  {\tt gm\_pa\_t}
  &  Represents a physical address. \\\hline
  {\tt gm\_region\_t}
  &  Represents an allocated range of virtual addresses within a virtual address
    space. \\\hline
  {\tt gm\_prot\_t}
  &  Defines the permitted types of access (such as read or write) to a
    virtual address. \\\hline
  {\tt gm\_mapping\_set\_t}
  &  A collection of mapped regions within a virtual address space that the
    device driver regards as logically related.   \\\hline
  {\tt gm\_as\_alloc\_t}
  &  Defines the composable flags for setting the virtual address allocation
    policy within a virtual address space. \\\hline
  {\tt gm\_mmu\_t}
  &  Encapsulates the operations for creating, modifying, and destroying mappings
    for a specific kind of MMU.  Also includes the page sizes supported
    by the MMU. \\\hline
  {\tt gm\_dev\_t}
  &  Represents a device, including the type of MMU ({\tt gm\_mmu\_t}) that the
    device uses.  A predefined, special instance of this type,
    {\em gm\_dev\_CPU}, represents the host CPU(s). \\\hline
  {\tt gm\_mmu\_mode\_t}
  &  Defines the flags controlling whether the device will use a shared or
    coherent page table. 
    If sharing is specified, but no other attached devices have the 
    same page table format, then a new coherent page table will be created. \\\hline
  {\tt gm\_dev\_cap\_t}
  &  Defines the composable flags for describing a device's capabilities,
    such as the ability to recover after an access that causes an address
    translation fault. \\\hline
  {\tt gm\_ret\_t}
  &  Defines the different return values from the public functions. \\\hline
  {\tt gm\_region\_placement\_t}
  &  
  Defines the physical memory placement/migration policy. \\\hline
  \end{tabular}
  }
  \caption{\label{tab:datatypes}Public data types defined by GMEM.}
\end{table*}

\paragraph{Virtual Address Space KPI.}

GMEM supports the following operations on virtual address spaces: 
\begin{itemize}
\item {\tt gm\_ret\_t gm\_as\_create(gm\_va\_t begin, gm\_va\_t end,
  gm\_as\_alloc\_t policy, gm\_as\_t **new\_as)}
  Creates a virtual address space supporting the specified virtual address
  range.  Virtual address allocation will be performed according to the
  specified policy, which is a composition of options like {\em first fit},
  {\em guarded} (which reserves an unused page on both sides of an region to
  trap out of bounds accesses), and {\em cached} (which enables an object cache
  of allocated but unmapped idle regions).
\item {\tt gm\_ret\_t gm\_as\_destroy(gm\_as\_t *as)}
  Destroys the virtual address space.
\item {\tt gm\_ret\_t gm\_as\_attach(gm\_as\_t *as, gm\_dev\_t *dev,
  gm\_mmu\_mode\_t mode, bool activate)}
  Attaches the specified device to the virtual address space using the
  specified mode (\textit{shared} or \textit{coherent}) of page table maintenance.  
  A device can be attached to
  multiple virtual address spaces, but must only be active within one at a
  time\footnote{The virtual functions of an SR-IOV device can be attached and
  active in different virtual address spaces simultaneously}.  If
  {\tt activate} is true, then the device is activated within the virtual
  address space.
\item {\tt gm\_ret\_t gm\_as\_alloc(gm\_as\_t *as, gm\_va\_t hint,
  gm\_va\_t size, gm\_va\_t align, gm\_va\_t no\_cross, gm\_va\_t max\_va,
  gm\_region\_t **new\_region)}
  Allocates a range of virtual addresses of the specified size, and satisfying
  the given constraints, such as alignment, from the virtual address space.
  Depending on the virtual address space's allocation policy, the given hint
  may be used to direct the search for free space.  The allocated region maintains
  a reference to its containing address space.
\item {\tt gm\_ret\_t gm\_as\_lookup(gm\_as\_t *as, gm\_va\_t addr,
  gm\_region\_t **out\_region)}
  Searches the virtual address space for an allocated region containing the
  specified virtual address.
\item {\tt gm\_ret\_t gm\_as\_synchronize(gm\_as\_t *as)}
  Waits for any pending, asynchronous mapping changes within the virtual
  address space to complete.
\end{itemize}
While mappings may be private to a specific device, depending on the mode
specified during device attachment, all attached devices share a single,
consistent view of the allocated regions within a virtual address space.

\paragraph{Device KPI.}

GMEM supports the following operations on devices: 
\begin{itemize}
\item {\tt gm\_ret\_t gm\_dev\_create(gm\_mmu\_t *mmu, void *mmu\_data,
  gm\_dev\_cap\_t cap, gm\_dev\_t **new\_dev)}
  Creates a representation for the device.  {\tt mmu\_data} is an opaque
  pointer that is stored within the device representation and passed to all
  {\tt mmu} operations.  In particular, this pointer is used with the
  {\tt gm\_mmu\_CPU} device to identify a virtual address space from
  the host CPU whose mappings should be inherited by a {\tt gm\_as\_t} upon
  attachment.
\item {\tt gm\_ret\_t gm\_dev\_destroy(gm\_dev\_t *dev)}
  Destroys the device.
\item {\tt gm\_ret\_t gm\_dev\_switch(gm\_dev\_t *dev, gm\_as\_t *as)}
  Activates the device within the specified virtual address space, implicitly
  deactivating the device elsewhere.
\item {\tt gm\_ret\_t gm\_dev\_detach(gm\_dev\_t *dev, gm\_as\_t *as)}
  Detaches the device from the specified address space, deallocating any
  device-specific mapping structures, such as a page table.
\item {\tt gm\_ret\_t gm\_dev\_fault(gm\_dev\_t *dev, gm\_va\_t addr,
  gm\_prot\_t access)}
  Invoked by a device driver when an address translation fault arises.  If
  the device is capable of recovering from address translation faults and
  the function is able to establish a valid mapping for the faulting address,
  then the function's return value will indicate to the driver that it should
  instruct the device to retry the faulting access. 
\item {\tt gm\_ret\_t gm\_dev\_register\_physmem(gm\_dev\_t *dev, gm\_pa\_t begin,
  gm\_pa\_t end)}
  Invoked by a device driver to register device local physical memory.
\end{itemize}
The exact behavior of coherent mode attachment depends on whether the
device is declared as capable of recovering from address translation faults.
Specifically, if a device cannot recover, then the only mappings created by
other devices that are replicated to the device are those that were
explicitly wired by their creator.  As part of GMEM's integration with the
host VM system, if the host CPU and a device are attached to a virtual address
space in coherent mode, then the wiring of a region by the host CPU through
host VM system interfaces will automatically replicate the mapping to the
device.

\paragraph{Region and Mapping Set KPI.}

GMEM supports the following operations on regions and mapping sets: 
\begin{itemize}
\item {\tt gm\_ret\_t gm\_region\_map(gm\_region\_t *region, gm\_prot\_t prot,
  struct vm\_page *pages[], gm\_mapping\_set\_t *set,
  int flags, void (*async\_callback)(void *), void *callback\_arg)}
  Creates a wired mapping from the region to the specified physical pages.  If
  {\tt set} is not NULL, then the mapping will be added to that set.  The given
  flags optionally specify the page size for the mapping and/or that the
  mapping be created asynchronously, in which case, the optional callback will
  be invoked when the mapping is complete.  (Alternatively, the GMEM user can
  call the synchronize function to wait for completion.)
\item {\tt gm\_ret\_t gm\_region\_unmap(gm\_region\_t *region,
  int flags, void (*async\_callback)(void *), void *callback\_arg)}
  Unmaps the region.  Like the map function, the given flags can specify
  asynchronous unmapping, with completion notification delivered through the
  callback or the synchronize function.  (Unmapping a region
  does not deallocate the region, it may be reused by a future mapping
  operation.)
\item {\tt gm\_ret\_t gm\_region\_dealloc(gm\_region\_t *region)}
  Frees the region from its virtual address space. If the region is still
  mapped upon deallocation, it will be unmapped.
\item {\tt gm\_ret\_t gm\_mapping\_set\_unmap(gm\_mapping\_set\_t *set,
  int flags, void (*async\_callback)(void *),
  void *callback\_arg)}
  Behaves similarly to the region unmap function, except this function
  unmaps the entire set of mappings.  Completion notification occurs when
  all of the mappings have been destroyed.
\item {\tt gm\_ret\_t gm\_region\_set\_policy(gm\_region\_t *region,
  gm\_dev\_t *dev, gm\_region\_placement\_t policy)}
  Sets the placement policy when a given region is faulted upon. 
  {\tt dev} determines where the physical memory should be allocated from upon
  a page fault if no physical mapping of the faulted address exists in
  the VA space. Specifically, if {\tt dev} is NULL, the physical
  memory will be allocated from the local physical memory of the
  faulting device. If the faulting device has no local physical memory,
  then allocate from host physical memory.


  Each region upon allocation has a default policy with {\tt dev} =
  NULL and {\tt policy} = {\tt UNIQUE}. {\tt UNIQUE} defines that
  there should always be a unique physical mapping of the faulted
  address in the VA space. If so ({\tt UNIQUE}), and any physical
  mappings exist in other separate page tables within the same VA
  space upon a page fault, all those physical mappings will be
  destroyed first.

\end{itemize}

\subsection{Use example \#1: pass-through device coordinating a KVM guest}
\label{sec:interface_exp_1}

The first example demonstrates using a pass-through device on a KVM
guest that dynamically changes its guest physical memory capacity.
This requires coordination between the EPT and the IOMMU; Linux's
low-level coordination mechanisms do not support this use case.

However, this coordination task can be easily offloaded to GMEM by using
the GMEM KPI to attach both the guest virtual machine
and the pass-through device to the same VA space.
KVM should first use
\texttt{gm\_as\_create} to create a VA space, representing the guest
physical to machine physical address mapping. Then, KVM should use
\texttt{gm\_dev\_create} to create a device that represents the guest
virtual machine's CPU, with EPT MMU operations. This device should then be
attached to the VA space with \texttt{gm\_as\_attach}.
After that, KVM should similarly call
\texttt{gm\_dev\_create} and \texttt{gm\_as\_attach} for the
pass-through device, registering IOMMU operations.

After attaching both devices to the same VA space, GMEM internally
handles the coordination when the guest physical memory is either
expanded or shrunk. KVM can use \texttt{gm\_as\_alloc} and
\texttt{gm\_region\_dealloc} to expand or shrink the size of guest
physical memory. During such a process, GMEM internally guarantees
that the pass-through device will consistently have a coherent view
of the available guest physical memory.

Note that if the pass-through device cannot retry and recover from a DMA
fault, KVM should additionally use \texttt{gm\_region\_map}
to wire the guest physical memory. If the device can recover
from  DMA faults, KVM should use \texttt{gm\_dev\_fault} to
resolve device page faults.

\subsection{Use example \#2: Discrete GPUs coordinating CPU}
\label{sec:interface_exp_2}

The second example demonstrates an OpenCL application that processes
buffers that are shared by both the CPU and a discrete GPU. The discrete GPU has
its local physical memory, a dedicated MMU, and can recover from GPU
page faults. This means that, without GMEM, the GPU driver must implement a full
VM management stack. The OpenCL application also demands a shared
virtual address space, which requires complete coordination between the two VM
systems. This cannot be handled by Linux's low-level
notification mechanisms.

However, this coordination can easily be achieved using GMEM.
When the application process is forked,
the core VM should use \texttt{gm\_as\_create}
and \texttt{gm\_as\_attach} to attach a GMEM device representing the
CPU (\texttt{gm\_dev\_CPU}) to the process's VA space. After that, the
GPU driver should similarly use \texttt{gm\_dev\_create} and
\texttt{gm\_as\_attach} to attach another GMEM device representing the
GPU to the same VA space. However, in this case, the GPU driver should
additionally call \texttt{gm\_dev\_register\_physmem} to register the
GPU's local physical memory. The GPU driver then resolves GPU page
faults by calling \texttt{gm\_dev\_fault}.

The OpenCL application can then benefit from a shared virtual address space.
For example, it can use the CPU to pre-process data in a buffer, then
directly issue a GPU kernel to process the data. GMEM transparently
handles the memory transfers between the CPU and GPU. Additionally,
the OpenCL application can transparently oversubscribe the GPU memory.
Upon GPU memory pressure, GMEM will swap out cold GPU data back to the
CPU DRAM. Advanced programmers who have insights into the data access
patterns can invoke \texttt{gm\_region\_set\_policy} to instruct GMEM
to switch between fault-driven migration or remote access on certain
VA regions. This allows application-specific
knowledge to easily be used to improve performance.

\section{Implementation Details}
\label{sec:impl}

GMEM is implemented in FreeBSD 13. This section describes
the differences between GMEM and a traditional core VM
system. Then it introduces a novel page table scheme for serving
disjoint requests and a device simulation infrastructure to help
evaluate GMEM's impact.

\subsection{Attaching peripherals to VA spaces}

To allow peripherals to be attached to processes' address spaces, GMEM
decouples FreeBSD's VA space management from CPU MMU functions.
Peripheral drivers must register MMU functions declared by GMEM,
including destroying a page table entry and invalidating a TLB entry.
These MMU functions will be invoked by GMEM to coordinate among
devices attached to each VA space.

FreeBSD's core VM system uses a binary search tree (splay tree) to
allocate and track virtual addresses. This is sufficient for both the
CPU and peripherals -- Linux and other device drivers also utilize
binary search trees. Additionally, GMEM integrates recent innovations
that equip the VA allocator with per-CPU caches to optimize
throughput of concurrent allocations~\cite{iommu2, vmem,
brett_thesis}.

\subsection{Integrating logical mappings with devices}

FreeBSD's logical mapping manager indicates whether the virtual
address should be backed by a disk file or zero-filled, so the
physical memory management system knows how to prepare the physical
pages. GMEM further extends such management by integrating device
local physical memory. Therefore, the logical mapping manager can
indicate whether a virtual address is backed by a device physical
page.

This enables GMEM to take charge of the device local physical memory.
For example, in a VA space shared by the CPU and a peripheral, the
program uses system calls like \texttt{mmap} to allocate anonymous
memory. The virtual address manager then allocates valid virtual
addresses and the logical mapping management system marks these addresses
as zero-filled. When the peripheral first faults on these
virtual addresses, device local physical pages are allocated,
zero-filled, and mapped. After that, if the CPU faults on these
virtual addresses again, the extended logical mapping manager will
indicate the virtual addresses have been backed by device physical
pages. GMEM can then decide whether to optimize data locality using
mechanisms like page migration.

\subsection{Maintaining coherent physical mappings}

GMEM internally coordinates page tables within the same address space
to provide a coherent view of memory. Two modes of coordination are
supported when devices are attached: shared and coherent.

Shared mode allows the attached devices to share the same page
table if they have compatible MMUs. Compatible MMUs can operate on
the same format of page table, but may use individual TLBs that
require TLB shootdowns upon page invalidation. GMEM implements such
TLB shootdowns with the TLB invalidation functions registered by
device drivers.

Coherent mode maintains an independent page table for the
attached device, which enables flexible data locality optimizations.
Note that coherent page tables need not contain identical mappings.
They simply need to provide coherent access to all memory in the
VA space at all times.
For example, when two devices attached to the same address space
access the same virtual address, GMEM can decide if the second access
should be remote or local. In the former case, GMEM replicates the
mapping in the page table of the second device. In the latter case,
GMEM can first tear down all mappings of the virtual page and then
migrate and map it to a local physical page of the second device.

\paragraph{Asynchronous MMU operations.}
GMEM additionally implements an asynchronous MMU operation queue for
each VA space, which helps alleviate the expensive MMU operation cost
for peripherals without compromising security~\cite{brett_thesis}.
Peripheral drivers can use GMEM KPIs to enqueue asynchronous MMU
operations with callback functions to be invoked at their completion
time. Additionally, they can use a GMEM KPI to wait for all pending
MMU operations. Besides, GMEM tries to batch unmapping requests by
coalescing TLB invalidation commands. All callback
functions of the coalesced requests will be invoked afterwards.

\subsection{Physical Memory Management}

GMEM allocates device local physical memory with a VMEM-based memory
allocator~\cite{vmem}. Peripheral drivers must first use the GMEM KPI to
register their local physical memory. GMEM then allocates the
corresponding kernel data structures and maintains three page queues
for actively used pages, free pages, and wired pages.

\paragraph{Device memory oversubscription.}

GMEM treats CPU physical memory as a swap space for device local
physical memory. This enables user applications to transparently
work on large datasets with GMEM-based drivers. Internally, GMEM may
perform memory migration in two directions when a device faults on CPU
memory under memory pressure. GMEM selects the most inactive device
physical pages and migrates them back to the CPU memory. After that,
GMEM migrates the faulted pages from the CPU to the device and resumes
the computation.

\paragraph{Bulk zeroing.}
In addition, GMEM integrates a bulk zeroing technique that
accelerates page zeroing speed and reduces the cost of page
faults~\cite{zhu2020comprehensive}. GMEM determines such a page
zeroing granularity based on the supported page sizes of the faulting
device.

\subsection{Reclaiming page tables with low contention}

Linux uses atomic operations to allocate page table pages and
exploits the fact that modifications to IOMMU page tables are
pairwise disjoint. However, this makes it difficult for Linux to
efficiently reclaim page table pages. GMEM includes a
reclaimable page table with low lock contention by adding reader-writer
locks. The reader lock is acquired when the mapping operation starts
to traverse the page table. The writer lock is only acquired when the
reference count of a page table page drops to zero during the
unmapping operation.

\subsection{Device simulation}

To better understand GMEM's impact on device driver development, a
device simulation infrastructure was built that constructs all the kernel
data structures that a real device driver would. Each device simulates an MMU
that translates virtual addresses for every memory access of the
device. Furthermore, a runtime library similar to OpenCL is
implemented to submit device computations by the user-level
application. The device computation is simulated by a kernel module
that utilizes the device MMU for every memory access. The kernel
module then converts the translated physical addresses to operate
on kernel direct map addresses.

Two heterogeneous computation workloads are simulated.
\texttt{VectorAdd} calculates the summation of two input vectors
prepared by the CPU to an output vector by the device. The output
vector is then read by the CPU again. \texttt{BP} trains a
three-layer fully-connected neural network with back-propagation. In
each training step, the CPU generates random input data and the
device uses them to train the neural network.

\section{Case Studies}
\label{sec:case-studies}

This section presents several case studies, representing the different
use cases from Table~\ref{table:va_cases} to demonstrate the impact of
GMEM on device driver development. The use case with a private VA
space and device local physical memory (such as CUDA) is not presented
here, as it is superseded by shared address spaces (such as UVM).
Both BUS\_DMA and OpenCL are evaluated using real devices with device
drivers rewritten to use the GMEM KPI.  An integrated GPU and a
discrete GPU are simulated with device drivers built from scratch.

\subsection{BUS\_DMA}

\begin{table*}[t]
\centering
\begin{tabular}{|l|l|l|l|l|l|l|l|l|}
\hline
Workload            & \#Con  & Linux    & FreeBSD  & GM-V     & GM-V-CAS & GM-V-RW  & GM-V-CAS-A & GM-V-RW-A \\ \hline
\multirow{3}{*}{TX} & 1      & 20.9(13) & 14.9(20) & 18.4(18) & 18.9(18) & 18.1(18) & 34.6(21)   & 28.7(18)  \\ \cline{2-9} 
                    & 4      & 37.6(45) & 20.8(47) & 27.5(52) & 29.6(62) & 27.4(41) & 37.7(29)   & 37.7(32)  \\ \cline{2-9} 
                    & 8      & 37.1(65) & 20.3(53) & 27.2(48) & 29.9(58) & 29.2(55) & 37.7(36)   & 37.7(47)  \\ \hline \hline
\multirow{3}{*}{RX} & 1      & 6.2(16)  & 5.8(13)  & 7.4(15)  & 8.3(16)  & 8.1(16)  & 12.2(18)   & 11.5(18)  \\ \cline{2-9} 
                    & 4      & 13.8(52) & 8.0(29)  & 10.5(26) & 12.7(33) & 11.4(28) & 23.9(40)   & 16.0(28)  \\ \cline{2-9} 
                    & 8      & 15.4(76) & 8.0(46)  & 12.4(44) & 13.7(50) & 13.1(45) & 23.7(52)   & 21.6(47)  \\ \hline \hline
RR                  & 8      & 63.8(28) & 60.1(6)  & 62.3(6)  & 61.3(6)  & 61.1(6)  & 61.0(6)    & 59.9(6)  \\ \hline
\end{tabular}
\caption{Netperf TCP performance with different number of connections. RX/TX report Gbps;
RR reports K transactions/sec. Numbers in brackets are CPU percentage.
The support machine uses Intel Xeon E3-1231 v3 CPU that has 4 physical 
cores with hyperthreading enabled.}
\label{table:iommu-perf}
\end{table*}


The BUS\_DMA module is a common kernel module that utilizes a
private VA space to provide security when sharing the CPU physical
memory. The private VA space is supported by an IOMMU driver that
allocates and manages I/O address mappings for DMA operations.
However, existing IOMMU drivers implement their own address mapping
systems that duplicate part of the virtual address management of the
core VM system. For example, FreeBSD's BUS\_DMA module spends 686
lines of code (LoC) implementing an I/O VA allocator with a red-black
tree.

There is a very high rate of churn for the memory mappings of an I/O
device, because buffers should only be mapped to I/O devices for the
brief period of the DMA operation. This provides ``strict" semantics
of security, as it prevents untrusted devices from having access to
memory that they do not need. However, the high churn rate of
mappings causes severe performance degradation from multiple layers
of the mapping system, including I/O VA allocation, page table
manipulations and invalidation of the IOTLB. To alleviate lock
contention, recent studies equipped the I/O VA allocator with per-CPU
caches~\cite{iommu2, brett_thesis}. Besides, Linux implemented a
lockless page table manipulation scheme, as the IOMMU driver only deals with disjoint mapping
requests. Furthermore, Gutstein implemented a novel asynchronous
unmapping memory mechanism that coalesces IOTLB invalidation without
compromising security~\cite{brett_thesis}. Note that none of these
innovations are coupled to any specific hardware characteristics.

The above innovations have been integrated with GMEM. After rewriting
FreeBSD's IOMMU driver with the GMEM KPI, the BUS\_DMA module can 
directly benefit from these innovations. This eliminates the
686-LoC I/O VA allocator, though it takes 53 extra LoC to utilize
GMEM KPI. We also improve Linux's lockless page table
scheme to allow page table page reclamation. This adds
another 78 LoC and replaces FreeBSD's 274-line giant-locked IOMMU page
table code. In total, the new driver added 131 LoC and
deleted 960 LoC.

The GMEM-based IOMMU driver enables a new BUS\_DMA API and higher
performance. Table~\ref{table:iommu-perf} shows the network
performance of an Intel 40Gbps NIC evaluated with netperf with
Linux's, FreeBSD's and FreeBSD's GMEM-based IOMMU drivers with the
same ``strict" security policy. Linux's NIC driver is patched to
enforce the strict security policy while FreeBSD's IOMMU driver is
patched to fix an algorithmic bug of the red-black tree used by its
I/O VA allocator. The GMEM-based driver has five variants. GM-V is a
GMEM-based IOMMU driver that enables a VMEM VA allocator. GM-V-CAS
uses Linux's lockless IOMMU page table scheme that cannot reclaim
page table page reclamation. GM-V-RW improves the page table scheme
by allowing page table page reclamation with the help of
reader-writer locks. In addition, GM-V-CAS-A and GM-V-RW-A utilizes
the asynchronous unmapping mechanism to coalesce IOTLB invalidation,
which requires the NIC driver to take advantages of the new BUS\_DMA
API.

GM-V-CAS utilizes similar optimizations as those in Linux's IOMMU
driver, so their network performance difference mostly comes from the
network stack. Linux's network stack has a built-in asynchronous DMA
mechanism in the transmit path, which brings 27\% higher transmit
(TX) throughput with lower CPU utilization compared with GM-V-CAS
based on FreeBSD. On the other side, FreeBSD's network stack does not
aggressively optimize for throughput, so GM-V-CAS obtains 11\% less
receive (RX) throughput consuming 34\% less CPU. It also allows
similar RR throughput with significantly lower (79\% less) CPU
compared with Linux.

GM-V-CAS-A and GM-V-RW-A further exploits the asynchronous unmapping
mechanism that batches the unmapping requests to reduce IOTLB waiting
time. Note that Linux's IOTLB batching mechanism does not provide the
same level of security, because it may expose a timing window that
the IOTLB still caches invalidated IOMMU mappings. GM-V-CAS-A
achieves similar line-rate TX throughput as Linux, but consumes 45\%
less CPU resources. On the receive path, it obtains 54\% higher
throughput while utilizing 32\% less CPU than Linux. GM-V-RW-A allows
reclaiming IOMMU page table pages compared with GM-V-CAS-A. It has
slightly higher CPU utilization on the TX path, and 9\% lower
throughput on the RX path. All GMEM-based variants do not noticeably
impact the RR performance.

\subsection{OpenCL}

OpenCL is a programming framework for CPU and hardware accelerators.
Our case study focuses on programming Intel's integrated GPU with
OpenCL's SVM feature, which utilizes a shared VA space to improve
user-level programmability. This is supported by Intel's GPU driver.
It implements its own coordination with the core VM system to share
the host physical memory. When the user program allocates an SVM
buffer with OpenCL API \texttt{clSVMAlloc}, the runtime library uses
system calls like \texttt{mmap} to allocate CPU virtual memory and
asks the GPU driver to track the buffer. Upon launching a GPU kernel,
the GPU driver pins all associated buffers on the CPU side -- it
scans the CPU page table, explicitly invokes CPU page fault handler
and pins the physical pages (to avoid handling GPU page
faults).\footnote{Intel's manual indicates the existence of support 
for GPU memory access replay, but the driver does not
implement any fault-and-retry mechanism.} After that, the driver
translates CPU physical mappings to logical mappings. Then, GPU
physical mappings are created with these logical mappings. At this
time, the integrated GPU can coherently access to these SVM buffers.
If the SVM buffer is latter deallocated, the GPU driver destroys the
GPU physical mappings, unpins the CPU pages and use system calls to
free the CPU virtual memory.

Intel's GPU driver is refactored by using GMEM's KPI. This eliminates
the driver code of the above coordination mechanism. 
Consequently, the user program is no longer required to use
\texttt{clSVMAlloc} to allocate coherent buffers --- any buffers
allocated via system calls can be coherently accessed by the GPU.
This allows easier code transplantation and further improves the
user-level programmability. Furthermore, GMEM is integrated with a
bulk zeroing technique that increases memory zeroing speed and
reduces the number of page faults~\cite{zhu2020comprehensive}. The
GMEM-based GPU driver can thus specify a page preparation granularity
and fault pages with faster speed. When launching an OpenCL GPU
kernel to operate on a 400MB uninitialized SVM buffer, the GPU kernel
launch time is 58 ms with the original driver. However, the
GMEM-based driver reduces this kernel launch time by 45\% to 32ms
when specifying a 2MB page preparation granularity.

\subsection{Simulated Devices}

Due to the difficulties of refactoring large, existing device drivers, two GPU
devices are simulated to explore using GMEM API to
implement device drivers from scratch. These two devices include a
faultable integrated GPU that shares the host physical memory and a
faultable discrete GPU that has its own local physical memory. The
former case demonstrates a similar case study as Intel's integrated
GPU, but it allows us to exercise GMEM KPI with a fault-and-retry
mechanism. Both drivers implement this fault-and-retry mechanism to 
handle GPU page faults.

Two workloads are simulated to verify the functionality of the device
drivers, including \texttt{VectorAdd} and \texttt{BP}.
\texttt{VectorAdd} initializes two random vectors and calculates the
summation of them in the third vector. \texttt{BP} runs
back-propagation algorithm on a three-layer neural network. Both
workloads utilize heterogeneous resources -- they process data on
both CPU and GPU. They are written with a programming model assuming
OpenCL's system-level SVM support, which uses a shared virtual
address space and allows accelerators to directly operate on buffers
allocated by system calls.


\subsection{Simulating a faultable integrated GPU}

Just like Intel's integrated GPU, the simulated integrated GPU allows
switching between two page table modes, including directly sharing
the CPU process's page table or using a coherent and independent GPU
page table. Therefore, two page table modes were implemented with GMEM
KPI in its GPU driver to exercise both hardware capacity. 
In addition, the fault-and-retry mechanism calls GMEM KPI
\texttt{gm\_dev\_fault} to handle GPU page faults. This further
eliminates the need of explicitly prefaulting and pinning pages
compared with the refactored Intel's GMEM-based GPU driver.


The GMEM-based driver of the faultable integrated GPU takes a total
of 182 LoC on memory management, including 125 LoC for MMU-specific
functions and merely 57 LoC on the device-independent part.

\subsection{Simulating a faultable discrete GPU}

The second simulated device is a faultable discrete GPU equipped with
local physical memory. This demonstrates a case study similar to
Nvidia's UVM case. However, this case study provides more advanced
programming model as OpenCL's system-level SVM support, which does
not require allocating memory with a specific memory allocation API.


The GMEM-based driver of the faultable discrete GPU takes a total of
193 LoC, including only 69 LoC on the device-independent part. This
number is slightly higher than that of the previous simulated device,
because it needs to use additional GMEM KPIs to register its local
physical memory and the page zeroing function.


\begin{table}[t]
\begin{tabular}{|l|l|l|l|}
\hline
Dev-DRAM  & dev-zero-fill & host-to-dev & dev-to-host \\ \hline
100MB & 91.6MB    & 412.7MB     & 406.5MB     \\ \hline
200MB & 91.6MB    & 137.5MB      & 91.6MB      \\ \hline
\end{tabular}
\caption{Memory traffic of training a 137.5MB neural network with different device local physical memory.}
\label{table:nn_training}
\end{table}

This case study also demonstrates support for memory oversubscription
from GMEM-based drivers. Table~\ref{table:nn_training} shows the
amount of memory zeroing and memory transfers when running the
\texttt{BP} workload with different local physical memory capacities.
When memory capacity is smaller than the workload size, the workload
can run seamlessly without any user-level changes. However, more
automatic data transfers are triggered by GMEM's internal
coordination mechanism.

\subsection{Discussion}

These case studies are not meant to be exhaustive.  Rather, they are
illustrative of the capability and flexibility of GMEM.  The device
driver for {\em any} current or future peripheral can utilize GMEM to
perform memory management for the device.  The case studies illustrate
the simplicity of doing so.  While it may be somewhat complex to
disentangle the ad-hoc memory management being performed by many
current device drivers, it should be possible to do so for the
writers/maintainers of such drivers. Our simulated devices have
a simplified hardware architecture, but their GMEM-based
drivers still demonstrate the ease of driver development on the
device-independent component of the driver. Therefore, it should be relatively
straight-forward to write new device drivers utilizing GMEM.

\section{Conclusion}
\label{sec:conclusion}

GMEM provides generalized and centralized memory management support
for current and future systems equipped with various hardware
accelerators.  The GMEM KPI obviates the need for device drivers to
replicate complex memory management code, and instead offload such
tasks to the OS, where they belong.  This enables devices to
automatically integrate current and future VM optimizations and
capabilities.  Furthermore, a device can easily share an address
space coherently with the CPU and other devices.
GMEM automatically coordinates the memory
management of all devices attached to each address space and enables
programs to execute seamlessly on any device regardless of local
memory capacity by transparently using CPU DRAM as swap space.

GMEM could further be enhanced by adding automatic data locality
optimizations and additional user-selectable policies. One could
add policies to enable asynchronous data prefetching, for example,
to the existing remote access and fault-driven migration policies.


\section*{Acknowledgments}
\label{section:ack}

This work was funded in part by National Science Foundation (NSF) Award CNS-2008857.

\bibliographystyle{plain}
\bibliography{refs}

\end{document}